\documentstyle[12pt]{ets2000}
\begin{document}
\def\btab{\begin{tabular}}
\def\etab{\end{tabular}}
\def\rfr#1{eq.(\ref{#1})}
\def\rfrs#1#2{eqs.(\ref{#1})-(\ref{#2})}
\def\Rfr#1{Eq.(\ref{#1})}
\def\Rfrs#1#2{Eqs.(\ref{#1})-(\ref{#2})}
\def\dfa{\derp{\mtc{R}}{a}}
\def\dfm{\derp{\mtc{R}}{\mtc{M}}}
\def\dfog{\derp{\mtc{R}}{\og}}
\def\dfi{\derp{\mtc{R}}{i}}
\def\dfe{\derp{\mtc{R}}{e}}
\def\dfo{\derp{\mtc{R}}{\O}}
\def\cu{\rp{2}{na}}
\def\cd{\rp{1-e^2}{na^{2} e}}
\def\ctr{\rp{(1-e^2)^{1/2}}{na^{2} e}}
\def\cq{\rp{1}{na^{2}(1-e^2)^{1/2}\si}}
\def\bb{\bibitem}
\def\eqi{\begin{equation}}
\def\eqf{\end{equation}}
\def\mtc#1{\mathcal{#1}}
\def\mc#1#2{\left[\matrix{#1 \cr #2\cr}\right]}
\def\mat#1#2#3#4{\left(\matrix{#1 & #2 \cr #3 & #4 \cr}\right)}
\def\ml#1#2{\ol{(\matrix{#1 & #2 \cr})}}
\def\mtc#1{\mathcal{#1}}
\def\a{\alpha}
\def\b{\beta}
\def\ga{\gamma}
\def\d{\delta}
\def\ve{\varepsilon}
\def\e{\epsilon}   
\def\z{\zeta}
\def\et{\eta}
\def\th{\theta}
\def\vth{\vartheta}
\def\io{\iota}
\def\k{\kappa}
\def\l{\lambda}
\def\m{\mu}
\def\n{\nu}
\def\x{\xi}
\def\o{\omicron}
\def\vp{\varpi}
\def\p{\pi}   
\def\r{\rho} 
\def\vr{\varrho}
\def\s{\sigma}
\def\vs{\varsigma}  
\def\t{\tau}
\def\u{\upsilon}
\def\f{\phi} 
\def\vf{\varphi}
\def\ch{\chi}
\def\ps{\psi} 
\def\og{\omega}
\def\G{\Gamma} 
\def\D{\Delta}
\def\TH{\Theta}
\def\L{\Lambda}
\def\X{\Xi}
\def\S{\Sigma} 
\def\P{\Pi} 
\def\U{\Upsilon}
\def\F{\Phi}
\def\PS{\Psi} 
\def\O{\Omega}
\def\co{\cos{\O}}
\def\so{\sin{\O}}
\def\cog{\cos{\og}}
\def\sog{\sin{\og}}
\def\ci{\cos{i}}
\def\si{\sin{i}}
\def\vass#1{\left\vert\ #1 \right\vert}
\def\rp#1#2{{#1\over#2}}
\def\ol#1{\overline{#1}}
\def\gr{general relativity}
\def\derp#1#2{\rp{\partial{#1}}{\partial{#2}}}
\def\dert#1#2{\rp{d{#1}}{d{#2}}}
\def\su#1{\stackrel{#1}}
\def\somma#1#2{\sum_{#1}^{#2}}
\def\ct#1{\cite{#1}}
\def\lb#1{\label{#1}}
\def\df{\buildrel def\over=}
\def\lg{LAGEOS}
\def\lgg{LAGEOS II}
\def\jgr{{J. Geophys. Res.}}
\def\asas{{Astron. Astrophys.}}
\def\celmec{{Celest. Mech.}}
\def\lt{Lense-Thirring\ }
\def\grc{gravitomagnetic}
\def\grm{gravitomagnetism}
\def\prc{precession}
\def\nd{node}
\def\pg{perigee}
\def\mal{mean anomaly}
\def\cl{clock}
\def\eft{effect}
\def\mash{Mashhoon}
\def\grn{Gronwald}  
\def\lic{Lichtenegger}
\def\cors#1{{\it {#1}}}
\def\gecc{G_{lpq}}
\def\fincl{F_{lmp}}
\def\gsue{\rp{\gecc}{e}\propto e^{\vass{q}-1}}
\def\rappe{\rp{1-e^2}{e}}
\def\rappee{\rp{\sqrt{1-e^2}}{e}}
\def\unoe{1-e^2}
\def\unoee{\sqrt{1-e^2}} 
\def\gme{GM_{\oplus}}
\def\cpl{C^{+}_{lmf}}
\def\dw{\r_{w}}
\def\ds{\r_{s}} 
\def\derf{\dert{\fincl}{i}}
\def\derg{\dert{\gecc}{e}}
\def\icombl{l-2p+q}
\def\icombb{l-2p}  
\def\doto{\dot\O}   
\def\dotog{\dot\og}   
\def\dotm{\dot\ma}
\def\ma{{\mtc{M}}}
\def\grp{\gecc\propto e^{\vass{q}}}
\title{Tidal satellite perturbations\\
and the Lense-Thirring effect}
\author{
Lorenzo Iorio$^{1)}$
\thanks{Fax: +390805442470, E-mail: iorio@ba.infn.it }
and Erricos C. Pavlis$^{2)}$}
\affil{
1) Dipartimento Interateneo di Fisica dell' Universit{\`{a}} di Bari, \\
Via Amendola 173 70126, Bari, Italy\\
2) Joint Center for Earth System Technology, JCET/UMBC\\
  NASA  Goddard Space Flight Center, Greenbelt, Maryland, U S A  20771-0001  }
\footnotetext{
Fax: +390805442470, E-mail: iorio@ba.infn.it }
\begin{abstract}
The tiny general relativistic \lt\ \eft\ can be measured by means of a
suitable combination of the orbital residuals of the nodes of \lg\ and \lgg\
and the \pg\ of \lgg. This observable
is
affected, among other factors,
by the Earth' s solid and ocean tides. They induce long-period orbital
perturbations that, over observational periods of few years, may alias
the detection of the \grc\ secular trend of interest. In this paper we
calculate explicitly the most relevant tidal perturbations
acting upon \lg s and assess their influence
on the detection of 
the \lt\ \eft. The present day level of knowledge of the solid and ocean
tides allow us to conclude that their influence on it
ranges from almost $4\%$ over 4 years to less than $2\%$ over 7 years. 
\end{abstract}
\section{Introduction}
In the past few years we have seen increasing efforts, both from a
theoretical and an experimental point of
view, devoted to
the measurement, by means of artificial satellites techniques, of
certain very small general relativistic effects
in the weak gravitational field
of the Earth. We refer to the small forces, unknown in classical Newtonian
mechanics,
induced by the rotation of the Earth itself on near orbiting bodies; due
to their formal
analogy with
the magnetic forces generated by electric charge currents they are
called
\grc\ \eft s.
Among them we recall
the dragging of the
orbital plane of a point mass and the shift of the pericenter' s
orbit in its plane, called \lt\
\eft\ (Lense and Thirring 1918;
Ciufolini and Wheeler 1995), which is presently  measured by
analyzing the orbits of \lg\ and
\lgg\ SLR satellites (Ciufolini 2000).
\par
In the \lg\ experiment the observable is a suitable
combination
of the orbital residuals of the nodes $\O$ of  \lg\ and 
\lgg\ and the \pg\ $\og$ of \lgg\ (Ciufolini 1996):
\eqi 
\d\dot\O^{I}+c_1\d\dot\O^{II}+c_2\d\dot\og^{II}\simeq 
60.2\m_{LT},\lb{ciufor}\eqf
in which $c_1=0.295$, $c_2=-0.35$ and $\m_{LT}$ is the parameter
which must be measured. It is 1 in \gr\ and 0 in
Newtonian mechanics.
General relativity predicts for the combined residuals a linear trend  with
a slope of $60.2$ mas/y.
Indeed:
\begin{eqnarray}
\dot \O_{LT}^{LAGEOS}&\simeq& 31 \ mas/y,\lb{letr1}\\ \dot
\O_{LT}^{LAGEOS
II}&\simeq& 31.5 \ mas/y, \\ \dot \og_{LT}^{LAGEOS II}&\simeq& -57\
mas/y.\lb{letr2}\end{eqnarray}
\par
In order to assess the feasibility of this space-based experiment the
error budget is of the utmost
importance, in particular with respect to the systematic uncertainties
induced by other physical
forces acting upon the combined residuals of \rfr{ciufor}.  Among them 
the orbital perturbations  induced
by the
Earth' s solid and
ocean tides (Dow 1988; Christodoulidis et al. 1988;
Casotto 1989) are very relevant. 
Ciufolini claims that \rfr{ciufor} is not affected by
the $l=2,4\ m=0$ part, both statical and dynamical, of the Earth' s
gravitational field. If confirmed, 
this feature is very important since the
systematic errors of the terrestrial gravitational field
which are most relevant for the detection of the \lt\ \eft\ lie just in the first
two even zonal harmonics (Ciufolini 1996).
\par
In order to model as more accurately as possible the
tidal perturbations on the \lg s and  to test 
quantitatively the independence of the proposed combined
residuals of the $l=2,4\ m=0$ harmonics a complete analysis of the
Earth tidal perturbations (Pavlis and Iorio 2000) on the
nodes of LAGEOS
and LAGEOS II
and the perigee of \lgg\ is carried out. The obtained results are used in
assessing the level at which the poor
knowledge
of the Earth' s solid and, in particular, ocean tides affects the
measurement of $\m_{LT}$ by means of \rfr{ciufor}.
\section{Earth' s solid and ocean tidal perturbations on \lg s' orbits}
The tidal perturbations are worked out in the framework of the linear
Lagrange
perturbation theory (Kaula 1966); since the \lt\ \eft\ is a linear
trend,
only the
long-period perturbations
averaged over an orbital revolution are considered.
\par 
The investigation of the tidal perturbations involves only
the terms of degree $l=2$ for the solid
tides
and also the terms
of degree $l=3,\ 4$ for the ocean tides. Concerning the latter ones,
only the prograde waves are considered.
\par
The
perturbative
amplitudes are inversely proportional to the perturbations'
frequencies. Concerning the
examined even degree perturbations, the periods depend only 
on the relative motion, as viewed in a geocentric inertial frame,
between the constituent' s tidal bulge and the orbital plane.
Concerning the odd degree ocean tidal perturbations, also the perigee' s
motion must be considered. So, in general, the spectrum of the
orbital tidal perturbations is richer than that of the tides as
viewed on the Earth' surface: often it happens that a constituent
which on the Earth gives rise to negligible tides induces orbital 
perturbations with relevant amplitudes and different periods.
\par
In general, as far as the solid and ocean tides'  constituents whose 
perturbations are examined here, it must be noted that, for $l=2$, the 
ocean perturbative amplitudes amount to at most $10\%$
of the corresponding solid perturbative amplitudes.
\par
The calculations are performed by using for the Love numbers $k_2$
the complex, frequency-dependent values released by (McCarthy 1996), for
the equilibrium tidal heights $H^{l}_{m}$ the values released by (Roosbeek
1996) and for the ocean tidal heights $C^{+}_{lmf}$ the
 coefficients of the
recent EGM96 model (Lemoine et al 1998).
\subsection{The solid tides}
Concerning the solid tides, the most effective 
constituents turn out to be the semisecular 18.6-year, the $K_1$ and the
$S_2$ which induce
on the examined \lg s' elements  perturbations 
of the order of $10^{2}-10^{3}$ mas with periods ranging from
111.24 days for the $S_2$ on \lgg\  to 6798.38 days for the
18.6-year tide.
\par
The latter, with its amplitudes of
-1079.38 mas, 1982.16 mas and -1375.58 mas for the nodes of the  \lg s and the
\pg\ of
\lgg\ respectively, could be
particularly insidious for
the detection of the \lt\ \eft. Indeed, since the observational periods 
adopted until now (Ciufolini 2000) range only from  3.1 to 4 years, it
could resemble a superimposed linear trend which may alias the recovery
of $\m_{LT}$. However, its effect should vanish since it is a $l=2,\ m=0$
tide, and \rfr{ciufor} should be not sensitive to such tides. This feature
will be quantitatively assessed later.
\par
Also the $K_1$ cannot be neglected: it induces on the \lg' s
node a perturbation with a period of 1043.67 days and an amplitude of
1744.38 
mas, while
the node and the \pg\ of \lgg\ are shifted by an amount of -398 mas and
1982.14 mas, respectively, with a period of -569.21 days.
\subsection{The ocean tides}
Regarding the ocean tides, whose  knowledge is less accurate
than that of the solid tides,
the $l=3$ part of the tidal spectrum turns out to be very interesting for
the
\pg\ of \lgg.
\par
Indeed, for this element the $K_1$ $l=3\ p=1,2\
q=-1,1$ terms
induce perturbations whose amplitudes and periods are
of the same order of magnitude of those generated by the solid tides.
For example,
the $l=3\ p=1\ q=-1$ harmonic has a period of 1851.9 days (5.09
years) and an
amplitude of 1136 mas, while the $l=3\ p=2\ q=1$ harmonic is less
effective:  its amplitude amounts to 346.6 mas and its  period to -336.28 days.
\par
It
should
be noted that, contrary to the
first two even degree zonal perturbations
which do not affect \rfr{ciufor}, the diurnal odd degree tidal
perturbations are not
canceled out by the combined residuals
of Ciufolini. So, over an observational period of few years the $K_1\ l=3\
p=1\ q=-1$
harmonic
may alias the \grc\ trend of interest to us.
\par
The even degree ocean tidal
perturbations are not particularly
relevant: they amount to some tens of  mas or less, with the
exception of
$K_1$ which perturbs the node of \lg\ and the \pg\ of \lgg\ at a
level of $10^2$ mas.
\section{The effect of the orbital tidal perturbations on the
measurement of the \lt\ \eft}
For a given  observational period $T_{obs}$, the tidal perturbations can be 
divided between two main categories according to their periods $P$: those with
$P<T_{obs}$ and those with $P>T_{obs}$. The former ones, even if their
mismodeled amplitude is great so that they heavily affect the orbital
residuals, are not particularly insidious because their effect averages out
if $T_{obs}=nP,\ n=1,2...$. Conversely, the latter ones must be carefully accounted for
since they may alias the determination of
$\m_{LT}$
acting as superimposed bias. This is particularly true for those diurnal
and semidiurnal tides which should affect the combined residuals, like the
$K_1\ l=3\ m=1\ p=1\ q=-1$.
\par
A preliminary analysis has been performed by evaluating the left-hand side of
 \rfr{ciufor} by means of 
 the nominal tidal perturbative amplitudes worked out in the previous
sections for $T_{obs}$ =1 year. The calculations have been repeated also 
by means of the mismodeled amplitudes. They show that, not only the $l=2,4\ m=0$
tides tend to cancel out, but also that this feature extends to the $l=3\
m=0$ ocean tides.
\par
A more refined procedure will be described below for the diurnal and semidiurnal tides.
\subsection{Case a: $P<T_{obs}$}
The effect of this class of perturbations has been assessed as follows.
\par
The orbital residuals curve has been simulated with MATLAB by including 
the \lt\ trend as predicted by \gr, the main mismodeled tidal perturbations:
\eqi \d A_{f}\sin{(\rp{2\p}{P_{f}}t+\f_{f})},\eqf where $f$ denotes
the chosen harmonic,
and a
noise.
About the mismodeling $\d A$ we have assumed that the main source of
uncertainties are the Love number $k_2$ (McCarthy 1996) for the solid
tides,
the load Love numbers (Farrell 1972; Pagiatakis 1990) and the ocean tidal heights
$C^{+}_{lmf}$ (Lemoine et
al. 1998) for the ocean tides.
It has been decided to build into the MATLAB routine the capability 
to vary the time
series length $T_{obs}$, the time span $\D t$, the amplitude of the
noise,
and the
initial phases $\f$ of the harmonics. \par
The so
obtained simulated curves, for different choices of $T_{obs}$, have been
subsequently fitted with a least-square
routine in order to recover, among the other things, the parameter $\m_{LT}$.
This procedure has been repeated with and without the whole set of mismodeled
tidal signals so to obtain an estimate
$\D\m_{tides}=\m_{LT}(all\ tides)-\m_{LT}(no\ tides)$ of their influence on
the measurement of $\m_{LT}$. This analysis show that
$2\%<\D\m_{tides}<4\%$
for
$T_{obs}$ ranging from 4 years to 7 years.
\par The decision of including the whole set of tides has been motivated by the
fact that their least-square fitted parameters show that some of them are strongly correlated
with the \lt\ trend, in particular for 4 years. For longer time spans, as it
could be expected, the \grc\ trend emerges clearly against the background of the
tidal noise and also the single harmonics, describing a large number of complete
cycles, tend to mutually decorrelate.
\subsection{Case b: $P>T_{obs}$}
The effect of this class of perturbations has been assessed by means of different
approaches.\par
First, in a very conservative way, the averaged
value of
the mismodeled tidal signal under consideration over different $T_{obs}$ has been considered.
The analysis has been performed for the 18.6-year tide and the $K_1\ l=3\
p=1\ q=-1$.
It has been assumed on the
semisecular 
tide a mismodeling level of $1.5\%$  
due to the uncertainty at its frequency in the anelastic behavior of
the Earth' s mantle accounted for by the Love number $k_2$.
The ocean constituent has been considered unknown at a $6\%$ level 
due to the uncertainties in the load Love number of degree $l=3$ 
and in the tidal height coefficient $C^{+}_{lmf}$ as released by EGM96.
Over different $T_{obs}$ the latter
affects the determination of the \lt\ effect at a $2.3\%$ level   at most,
while the
zonal tide, as predicted by Ciufolini (1996), almost cancels out giving rise
to negligible contributions to the combined residuals.
\par
These results have been confirmed fairly well by adopting another method
analogous to that employed in (Vespe 1999) in assessing the influence of the 
direct solar radiation pressure' s eclipses on the determination of the
\lt\ \eft. We take the signal of interest over a given $T_{obs}$,
fit it with a straight line and compare the so obtained slope, in unit of 60.2
mas/y,  to $\m_{LT}$.
\acknowledgments{L. Iorio warmly thanks I. Ciufolini and L. Guerriero.
E. C. Pavlis gratefully acknowledges partial support for this project
from NASA's Cooperative Agreement NCC 5-339.}

\end{document}